\documentclass[a4paper,11pt]{article}
\pdfoutput=1 

\usepackage{jheppub} 

\usepackage{jheppub} 

\usepackage{comment}
\usepackage{enumerate}
\usepackage{physics}
\usepackage{color}
\usepackage{lineno}

\title{Axion dark matter searches from the standard halo over the tidal stream to the big flow}
\author[a]{Andrew K. Yi}
\author[b,1]{and Byeong Rok Ko\note{Corresponding author.}}
\affiliation[a]{SLAC National Accelerator Laboratory, 2575 Sand Hill Rd., Menlo Park, California 94025, USA}
\affiliation[b]{Department of Accelerator Science, Korea University Sejong Campus, Sejong 30019, Republic of Korea}

\emailAdd{andrewyi@slac.stanford.edu}
\emailAdd{brko@korea.ac.kr}

\abstract{
  The sensitivity of axion dark matter searches depends on the signal
  window that results from the velocity dispersion of axion dark
  matter. Since the ratio of signal windows is about 6500 between the
  standard halo and the big flow axion dark matter, each axion dark
  matter search usually uses a separate data acquisition (DAQ) channel
  with a different frequency resolution bandwidth (RBW).
  In this work, we demonstrate axion dark matter searches covering the
  standard halo, the tidal stream, and the big flow employing a DAQ
  channel starting with a single high resolution RBW, without
  sacrificing the DAQ efficiency, where the DAQ process includes
  online fast Fourier transforms and writing the outputs to disk.  
  Assuming the total amount of data is sensitive to
  Dine-Fischler-Srednicki-Zhitnitskii (DFSZ) axion dark matter that
  follows the standard halo model and makes up 100\% of the local dark
  matter density, the same data can also be used for the tidal stream
  and the big flow axion dark matter searches that would be sensitive
  to DFSZ axion dark matter that constitute 19.2\% and 12.4\% of the
  local dark matter densities, respectively, at a 90\% confidence
  level.  
  We also report that the filtering of the individual power spectra
  acquired with a relatively high resolution RBW e.g., for the big
  flow search can prevent a possible significant degradation in the
  signal to noise ratio from the searches in the lower resolution
  RBW's, i.e., the standard halo and tidal stream searches.
}

\keywords{Dark Matter, Axions and ALPs, Beyond Standard Model}

\begin{document}
	
\maketitle
\flushbottom
	
\section{Introduction}
Cold dark matter (CDM) is believed to constitute about 85\% of the
total matter in our Universe from the standard model of Big Bang
cosmology and high precision cosmological measurements~\cite{PLANCK}.
The Standard Model of particle physics (SM) cannot explain CDM and the
characteristics of CDM are still veiled to date even though there is
strong evidence of its
existence~\cite{CDM-EVIDENCE1,CDM-EVIDENCE2}. One of the strong
potential CDM candidates is the axion~\cite{AXION1,AXION2} which was
originally invented as a natural solution of the strong $CP$ problem
in the SM~\cite{strongCP1,strongCP2,strongCP3,strongCP4,strongCP5} and
results from a breakdown of a new symmetry proposed by Peccei and
Quinn~\cite{PQ}. Considering the current galaxy formation, dark matter
has to be cold as aforementioned, or equivalently, non-relativistic,
massive, and stable, which all fit the characteristics of the
axion. Hence, axions are strong candidate which can be called ``axion dark matter.''

The axion haloscope search which utilizes axion-photon coupling and a
microwave resonant cavity proposed by Sikivie is the most sensitive
axion detection method in the microwave region as of today thanks to
the resonant conversion of axions to
photons~\cite{sikivie-PRL,sikivie-PRD}.
Recent axion haloscope search experiments realized the
Dine-Fischler-Srednicki-Zhitnitskii (DFSZ) axion~\cite{DFSZ1,DFSZ2}
sensitivity~\cite{ADMX-DFSZ1,ADMX-DFSZ2,ADMX-DFSZ3,12TB-PRL,12TB-PRX},
assuming axions make up 100\% of the local dark matter density
$\rho_a$, i.e., $\rho_a=0.45$~GeV/cm$^{3}$,\footnote{The axion
community tends to use 0.45 GeV/cm$^{3}$ as a convention that was
established long before, and in order to compare the sensitivity among
different experiments it has remained that way. As dark matter density
is a scalable variable to reach results, it is possible to convert our
results here for given dark matter densities~\cite{DMRHO1,DMRHO2}.}
and axion dark matter follows the standard halo model
(SHM)~\cite{SHM}. The DFSZ axion can be implemented in grand unified
theories (GUT)~\cite{GUT}, thus supporting GUT as well as providing us
with the understanding of the total matter in our Universe if DFSZ
cold axion dark matter turns out to be responsible for 100\% of the
local dark matter density.

In addition to the standard halo dark matter, dark matter streams
complementary to the standard halo dark matter would contribute to the
local dark matter density in their
presence~\cite{TIDAL1,TIDAL2,BIGFLOW}.
Two well-known dark matter streams that have distinct velocities $v$'s
and velocity dispersion $\delta v$'s are dark matter of a tidal stream
from the Sagittarius dwarf galaxy~\cite{TIDAL1,TIDAL2} and that of the
``big flow''~\cite{BIGFLOW}.
Although our work here assumes the big flow exists based on
ref.~\cite{BIGFLOW}, we also point out that several
studies~\cite{BIG1,BIG2,BIG3} are skeptical of the feature proposed in
ref.~\cite{BIGFLOW}. The contention regarding the astrophysical
existence of the big flow itself is beyond this work's scope, as it is
more focused on optimizing the signal to noise ratio (SNR) across
different possible axion dark matter models of varying bandwidths
should they exist.
Advancements in various aspects of the haloscope experiment will lead
to better sensitivity, and this applies to all axion dark matter
models. Expected SNR values in this work reflect the current reach of
limits for an axion haloscope capable of searching for DFSZ axion dark
matter for the SHM.

The tidal stream would have a $v\simeq 300$~km/s and a
$\delta v\simeq20$~km/s, while the big flow a $v\simeq 480$~km/s and a
$\delta v\lesssim 53$~m/s, respectively.
For dark matter whose mass is around 1~GHz, such $v$'s and
$\delta v$'s result in the signal windows of about 150~Hz for the
tidal stream and 0.625~Hz for the big flow, respectively, which are
much narrower than that of the SHM dark matter of about 4~kHz. Because
of such big differences in the signal windows, e.g., $\sim$6500
between the SHM and the big flow, the required frequency resolution
bandwidth (RBW) $\Delta f$ of data for each axion dark matter search
is also quite different, which can be handled with separated data
acquisition (DAQ) channels with relevant
RBW's~\cite{ADMX-HR1,ADMX-HR2,ADMX-HR3,ADMX-HR4}. The difference
between the SHM and the tidal stream is relatively smaller, thus one
can perform both the SHM and the tidal stream axion dark matter
searches using a DAQ channel with a single appropriate RBW as done in
refs.~\cite{12TB-PRL,12TB-PRD}.
On the other hand, the recent interferometer based experiment also
showed a feasibility that can do virialized and non-virialized axion
searches in parallel~\cite{DALI-PRD}.

In this work, we demonstrate axion dark matter searches covering the
SHM, the tidal stream, and the big flow employing a DAQ channel with
only one high resolution (HR) RBW, without sacrificing the DAQ
efficiency $\epsilon_{\rm DAQ}$. Here the DAQ process includes online
fast Fourier transforms (FFTs) and writing the outputs to disk.
Using the data sensitive to the standard halo DFSZ axion dark matter
with $\rho_a=0.45$~GeV/cm$^3$, the tidal stream and the big flow axion
dark matter searches would be sensitive to DFSZ axion dark matter that
constitute 19.2\% and 12.4\% of the local dark matter densities,
respectively, at a 90\% confidence level (CL).
We also report that the filtering of the individual power spectra
acquired with an HR RBW e.g., for the big flow search can
prevent a possible significant degradation in SNR from the searches in
the lower resolution RBW's, i.e., the standard halo and tidal stream
searches. We are referring to the filtering for the big flow
searches~\cite{ADMX-HR1,ADMX-HR2,ADMX-HR3,ADMX-HR4} as
``HR-RBW-filtering'' in this work, where the HR RBW was chosen to be
0.025~Hz as explained in section~\ref{SEC:PARM}.
\section{Parameters}\label{SEC:PARM}
We used the parameters in ref.~\cite{12TB-PRL}. Particular parameters
relevant to this work are listed in table~\ref{parameters}.
The 13 power spectra with $\Delta f=0.025$~Hz correspond to 5200 power
spectra with $\Delta f=10$~Hz, and thus the expected sensitivity in this
work is similar to that reported in ref.~\cite{12TB-PRL}.
For dark matter whose mass is around 1~GHz, the expected signal window
of the big flow axion dark matter is about 0.625~Hz~\cite{BIGFLOW} as
mentioned earlier, corresponding to 25 frequency intervals with a
$\Delta f=0.025$~Hz.
By 25 coadding such frequency points, one can
search for the big flow axion dark matter. Other axion dark matter
models with wider signal windows can be done as well with relevant
data processing.
The choice of an intermediate frequency (IF) of 1~MHz avoids the
$\epsilon_{\rm DAQ}$ decreasing too much and also maintains a
manageable data size, which is to be discussed in
section~\ref{SEC:DAQ}.
\begin{table} [h]
  \centering
  \begin{tabular}{ | l | c | c | } \hline
               & ref.~\cite{12TB-PRL}     & this work\\ \hline\hline  
    frequency resolution bandwidth $\Delta f$ & 10~Hz    & 0.025~Hz \\ \hline  
    intermediate frequency (IF)               & 10.7~MHz & 1~MHz    \\ \hline  
    number of power spectra                   & 5120     & 13       \\ \hline  
    individual spectrum span                  & 150~kHz  & 150~kHz  \\ \hline  
    frequency tuning steps                    & 10~kHz   & 10~kHz   \\ \hline  
  \end{tabular}
  \caption{Parameters used in ref.~\cite{12TB-PRL} and this work. The
    numbers 5120 and 13 are numbers of power spectra taken for each
    resonance frequency step, where the former is in fact 40 power
    spectra with each being the average of 128 individual
    spectra~\cite{JINST-DAQ}.}  
  \label{parameters}
\end{table}
\section{DAQ efficiency}\label{SEC:DAQ}
We used the same DAQ framework developed in our previous
work~\cite{JINST-DAQ}, but used a different fast digitizer,
M4i-4480-x8 from Spectrum Instrumentation GmbH~\cite{SPECTRUM}. The
details of the DAQ realization are identical to those found in
ref.~\cite{JINST-DAQ}. The new fast digitizer also has two parallel
channels and the main specifications of the analogue-to-digital
converters (ADCs) are a 14-bit amplitude and a maximum sampling rate
of 400~MSamples/s. The relevant parameters for this work are listed in
table~\ref{DAQPARM}. The fast digitizer M4i-4470-x8 with a 16-bit
amplitude and a maximum sampling rate of 180~MSamples/s from Spectrum
Instrumentation GmbH~\cite{SPECTRUM} was used for
refs.~\cite{12TB-PRL, 12TB-PRD, JINST-DAQ}.
\begin{table} [h]
  \centering
  \begin{tabular}{ | c | c | c | } \hline
               & ref.~\cite{12TB-PRL}               & this work\\ \hline\hline  
    sampling rate (MSamples/s)                      & 45 & 3.125 \\ \hline  
    ADC data points per individual power spectrum   & 4.5M & 125M  \\ \hline  
  \end{tabular}
  \caption{Parameters for the DAQ process used in ref.~\cite{12TB-PRL} and this work.}  
  \label{DAQPARM}
\end{table}
Therefore, the 45~MSamples/s was chosen from (180~MSamples/s)/$2^2$ and
3.125~MSamples/s from (400~MSamples/s)/$2^7$.
\begin{figure}[h]
  \centering
  \includegraphics[width=0.95\columnwidth]{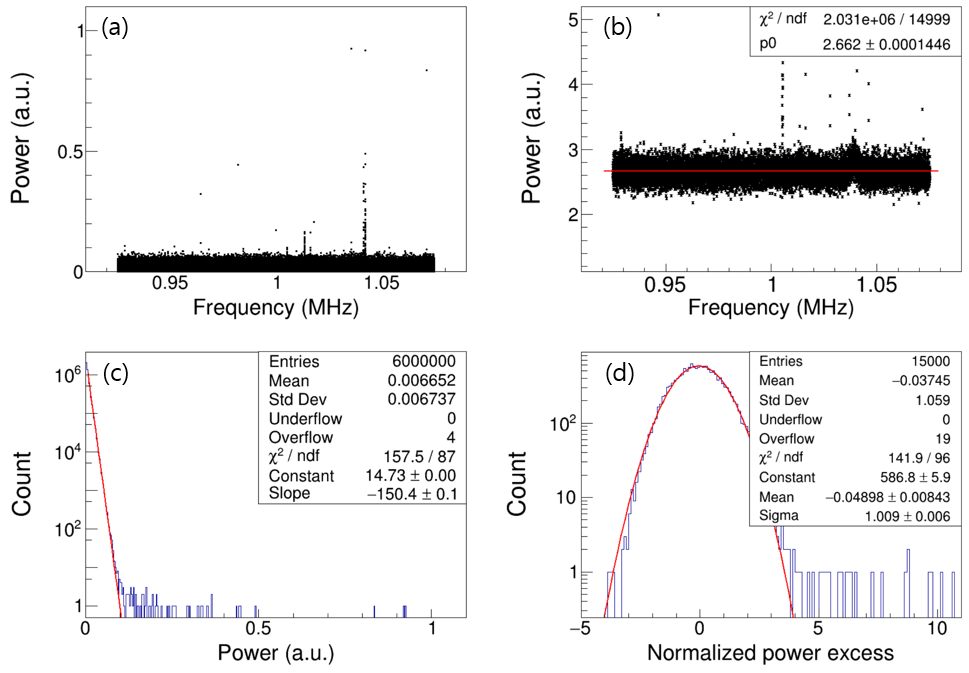}
  \caption{(a) shows a power spectrum in arbitrary units (a.u.) and
    $\Delta f=0.025$~Hz obtained from the digitizer, where typical
    narrow peaks are also detected on top of the random noise.    
    (b) results from (a) by merging 400 frequency points, and thus
    $\Delta f$ becomes 10~Hz.    
    (c) shows the power distribution from (a) and (d) the normalized power
    excess from (b) after background subtraction. The red lines in
    (b), (c), and (d) are a constant, an exponential, and a gaussian
    fit, respectively.}  
  \label{FIG:DATA}
\end{figure}
Figure~\ref{FIG:DATA} (a) shows a typical power spectrum
from the fast digitizer applying the parameters in table 2 and the
50-$\Omega$ terminations on the input ports as the noise sources. As
expected, the noise power distribution shown in figure~\ref{FIG:DATA}
(c) is almost exponential.
Merging 400 frequency points in figure~\ref{FIG:DATA} (a) results in
figure~\ref{FIG:DATA} (b) whose $\Delta f$ becomes 10~Hz and the
normalized power excess shown in figure~\ref{FIG:DATA} (d) follows a
gaussian distribution as expected. The data size of the individual
power spectra is about 210~MB to achieve $\Delta f=0.025$~Hz, which is
$\sim$2.7~GB for each resonance frequency tuning step taking 13 power
spectra, and $\sim$5.5~TB for $\sim$2000 resonance frequency tuning
steps~\cite{12TB-PRL}.
Retaining the IF of 10.7~MHz requires a sampling rate of at least
25~MSamples/s provided by the digitizer adopted in this work to
satisfy the Nyquist rate~\cite{NYQUIST}, hence resulting a data size
that is at least 8 times larger.
\begin{figure}[h]
  \centering
  \includegraphics[width=0.95\columnwidth]{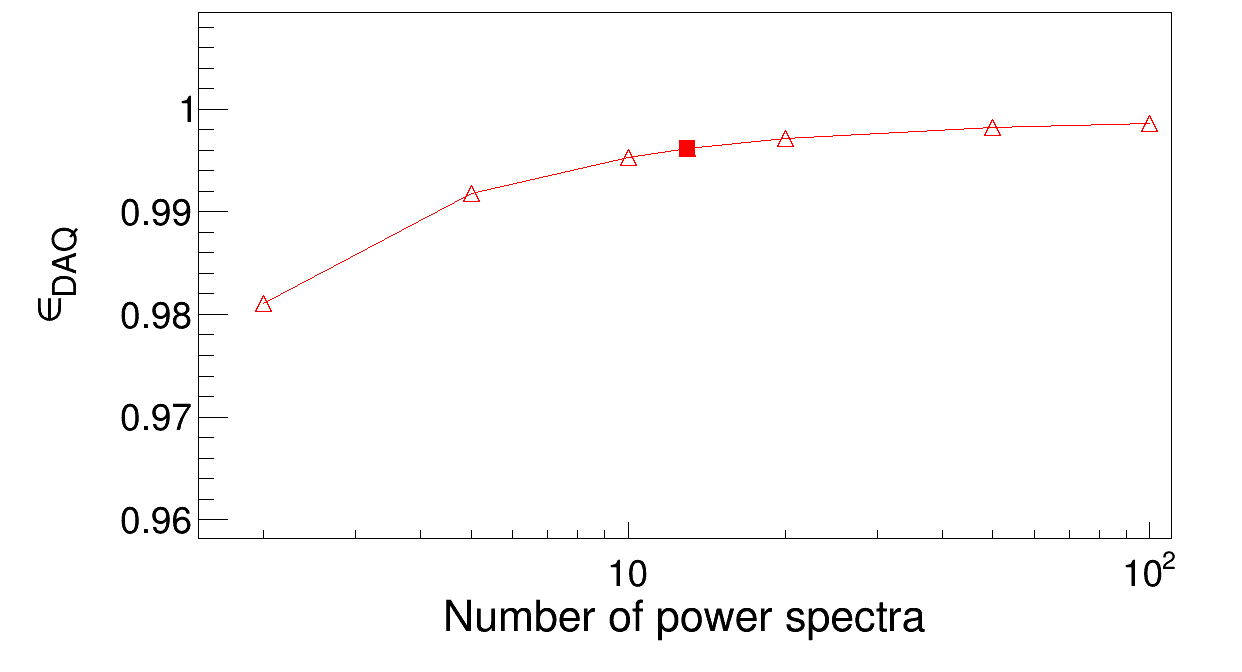}
  \caption{$\epsilon_{\rm DAQ}$ as a function of the number of power
    spectra. The case where 13 power spectra are taken is denoted by
    the solid rectangle.}
  \label{FIG:DAQEFFI}
\end{figure}

Our sampling rate of 3.125~MSamples/s introduces the signal aliasing
from frequencies higher than about 1.56~MHz according to the Nyquist
rate~\cite{NYQUIST} and this can be avoided using a commercial low
pass filter, e.g., LPF-B0R8+~\cite{MINICIRCUIT}. The filter works well
up to 1.1~MHz which is low enough to prevent the signal aliasing from
frequencies above $\sim$1.56~MHz.
The frequency down conversion to IF also introduces unwanted image
backgrounds which can be removed with an image rejection mixer. The
working IF range for the image rejection mixer
IRM0622B~\cite{POLYPHASE} used in refs.~\cite{12TB-PRL, 12TB-PRD} is
10$\sim$50~MHz which does not fit to the IF of 1~MHz for this
work. The IQ (in-phase and quadrature) mixer QD0622B~\cite{POLYPHASE}
is known to work for an IF value down to 100~kHz without introducing
the potential IQ imbalance, and the image backgrounds can be removed
with a proper signal processing combining I and Q signals from the IQ
mixer.\footnote{Note the IQ mixer and low pass filter were not
directly used in this work and they are suggestions to figure out the
addressed problems raised from the choice of such a low sampling
rate.} Therefore, the two parallel channels on the fast digitizer have
to be employed for the processing of the I and Q signals from the IQ
mixer.
The $\epsilon_{\rm DAQ}$ was then estimated in the same way used in
ref.~\cite{JINST-DAQ} only when the IQ mixer and the two parallel
channels were used, but the error estimation was done
differently. Since our $\epsilon_{\rm DAQ}$ does not follow the
binomial statistics,\footnote{We used binomial statistics in our
previous study~\cite{JINST-DAQ} which led to the overestimated
$\epsilon_{\rm DAQ}$ errors.} the $\epsilon_{\rm DAQ}$ error was
estimated by checking the DAQ processing time for every individual
power spectra.
The deviation was at an order of 0.01\% which is also consistent with
the variance associated with 250M ADC data points per individual power
spectrum. Therefore, such small error bars were not shown (were
ignored) in figure~\ref{FIG:DAQEFFI}.
The data acquisition time is 40~s with $\Delta f=0.025$~Hz and the
online FFT time for 250M ADC data points from the two parallel
channels is about 25~s. Hence, an extra 15~s is available and can be
allocated for the digitizer-to-memory transfer of the ADC data and
writing the FFT output to disk without any loss to
the $\epsilon_{\rm DAQ}$~\cite{JINST-DAQ}.\footnote{The ADC data was
not saved to disk.}
Because the next DAQ for the next resonance frequency step can
simultaneously start with the last FFT in the current DAQ process, the
time for the last FFT does not contribute to the
$\epsilon_{\rm DAQ}$~\cite{JINST-DAQ}.
For taking 13 power spectra, the $\epsilon_{\rm DAQ}$ is, therefore,
over 99\% and marked as the solid rectangle in
figure~\ref{FIG:DAQEFFI}.
Even for a higher IF,  the data acquisition time is 40~s for
$\Delta f=0.025$~Hz. However, if we were to take data for the higher
resolution of $\Delta f=0.025$~Hz for an IF of 10.7~MHz, online FFTs
for 2G ADC data points provided by a sampling rate of 25 MSamples/s
will need about 200~s. The expected $\epsilon_{\rm DAQ}$ for this case
is at most 20\% even ignoring the ADC and FFT data transfers. Such a
low $\epsilon_{\rm DAQ}$ is hardly accepted in general, and thus
lowering the IF from 10.7~MHz to 1~MHz is unavoidable.

\section{Axion dark matter searches}
\begin{figure}[h]
  \centering
  \includegraphics[width=0.95\columnwidth]{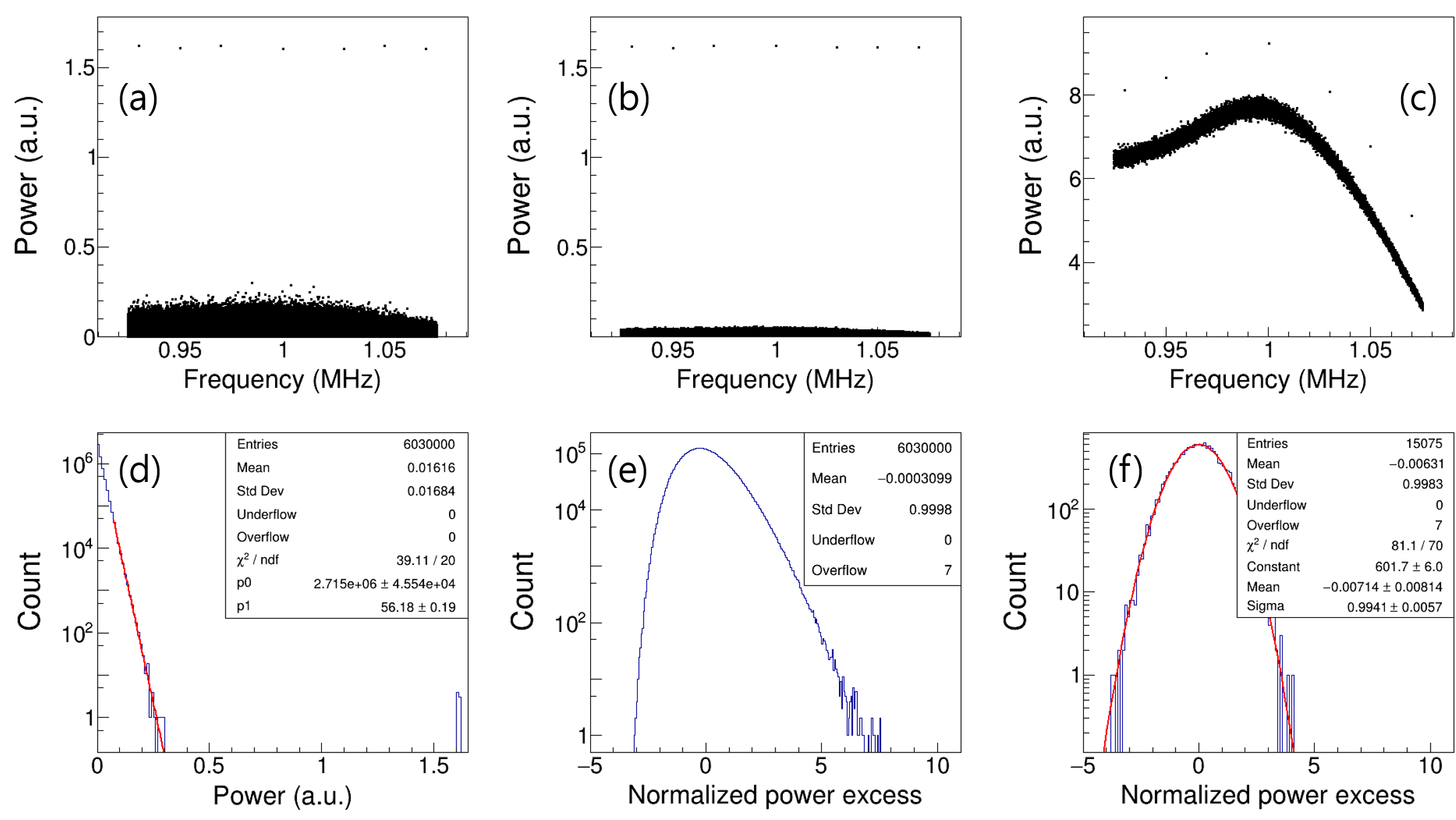}
  \caption{Simulation data when the axion mass matches the frequency
    of the cavity mode.
    (a) shows a simulated power spectrum with $\Delta f=0.025$~Hz,
    where 7 narrow peaks are also added and the 4th narrow peak is
    located at the signal region.
    (b) is an averaged power spectrum over 13 power spectra, where one
    of them is shown in (a).    
    (c) results from (b) by merging 400 frequency points, thus $\Delta
    f$ becomes 10~Hz.    
    (d) shows the power distribution from (a).
    (e) and (f) are the normalized power excess from (b) and (c) after
    background subtraction by a perfect fit, respectively.
    Note the 7 narrow peaks belong to the overflows in (e) and
    (f). The red lines in (d) and (f) are an exponential and a
    gaussian fit, respectively.}  
  \label{FIG:SIMUL}
\end{figure}

Unlike figure~\ref{FIG:DATA} (b), the background shape in typical
axion dark matter searches is not necessary flat, and looks like
figure~\ref{FIG:SIMUL} (c) due to complicated properties of the cavity
and the receiver chain.
One sample of the data for axion dark matter searches in this work is
shown in figure~\ref{FIG:SIMUL}, which is the case when the axion mass
matches the resonance frequency of the cavity.
Power at each frequency point whose frequency interval is 0.025~Hz
shown in figure~\ref{FIG:SIMUL} (a) was generated following a $\chi^2$
distribution with $2N_f$ degrees of freedom, where $N_f$ is the number
of frequency points~\cite{ADMX-HR2}.
As shown in figure~\ref{FIG:SIMUL} (a), (b), and (c), 7 narrow peaks
are embedded to check the effect of the HR-RBW-filtering to the SNR
efficiency $\epsilon_{\rm SNR}$.
The 7 peaks are located over the whole spectrum including the signal
region around the 4th peak.
One can add more such peaks, but the current number of peaks is
sufficient to represent the effects not only of the aforementioned
filtering but also of the cavity response for unwanted spurious
signals while the cavity is tuned to different frequency tuning
steps. The latter is seen in figure~\ref{FIG:SIMUL4} (b) later.
The HR-RBW-filtering condition for the data shown in
figure~\ref{FIG:SIMUL} removes frequency points with power levels
higher than 0.35. This value is obtained by identifying the outliers
of an exponential fit shown in figure~\ref{FIG:SIMUL} (d), and is
obvious through simple observation. Only the 7 unwanted narrow peaks
are filtered out and the data is treated like stationary Poisson
noise.
Since this filtering was applied to all the individual power spectra
whose RBW is 0.025~Hz, in this work it affects not only the big flow
search using individual power spectra in figure~\ref{FIG:SIMUL} (a),
but also the standard halo and tidal stream searches using that in
figure~\ref{FIG:SIMUL} (c).
The axion signal whose window size depends on the models, but with the
same mass is added to the background. Thus, three data samples are
used for the standard halo, the tidal stream, and the big flow,
respectively. We also have another set without the 7 narrow peaks and
refer to it as the ``reference sample.'' The $\epsilon_{\rm SNR}$ due
to background subtraction can be estimated analytically according to
ref.~\cite{ANALeSNR} and is expected to be at most 93\% using a
suitable Savitzky-Golay filter~\cite{SGFILTER}.
Typical $\chi^2$ parameterizations for background subtraction also
provided about the same efficiency, where the efficiency was estimated
from a large sample of simulation data~\cite{12TB-PRL}.
Because the background subtraction efficiency is not the interest of
this work, we used the simulation input background functions, i.e., a
perfect fit, for background subtraction. Therefore, applying a perfect
fit to the reference sample without any filtering of individual power
spectra results in the SNR with an $\epsilon_{\rm SNR}=100\%$, where
one filtering is the aforementioned HR-RBW-filtering for the big
flow searches~\cite{ADMX-HR1,ADMX-HR2,ADMX-HR3,ADMX-HR4} and the
other~\cite{HAYSTAC-PRD} for the standard halo and tidal stream axion
dark matter searches.
The latter filtering is referred to as the ``LR-RBW-filtering'' with
an RBW of 10~Hz in this work, where LR stands for low resolution. The
typical condition of the LR-RBW-filtering is for normalized power
excess points that are larger than 4.5. This also removes the 7 narrow
peaks corresponding to the 7 overflows shown in figure~\ref{FIG:SIMUL}
(f).
\begin{figure}[h]
  \centering
  \includegraphics[width=0.95\columnwidth]{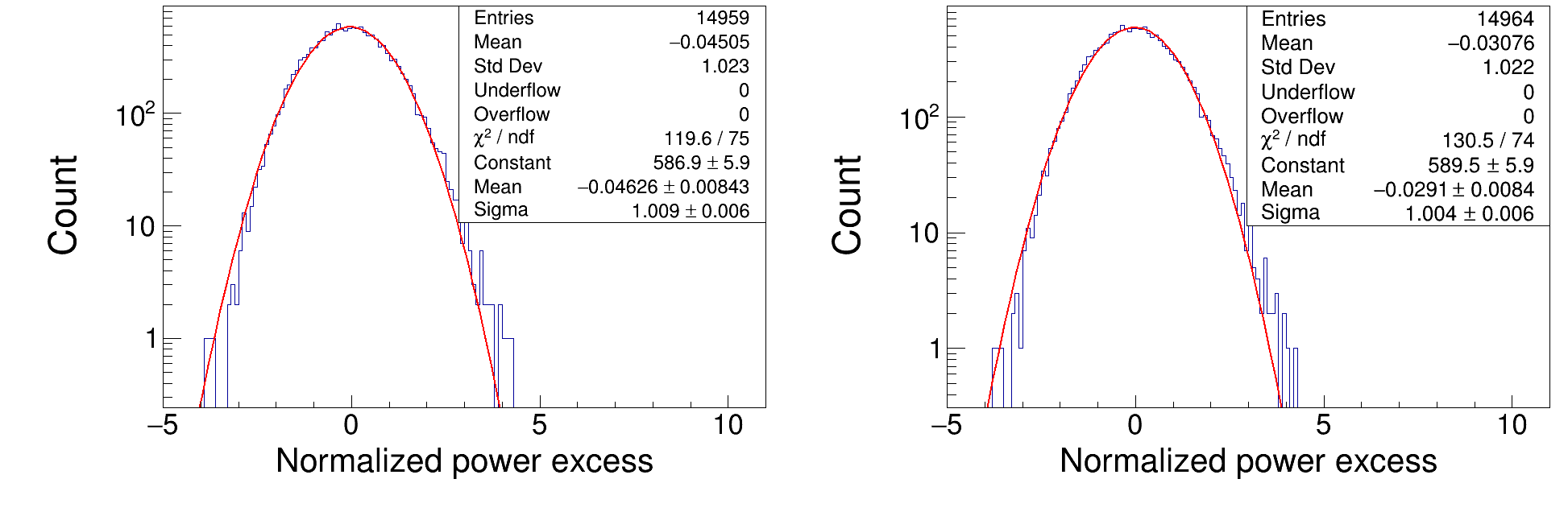}
  \caption{Normalized power excess distributions after applying the
    LR-RBW-filtering only (left) and both the HR-RBW-filtering and
    LR-RBW-filtering (right) to data shown in figure~\ref{FIG:DATA},
    respectively.}  
  \label{FIG:DATAFILTER}
\end{figure}

We applied the filtering to available data shown in
figure~\ref{FIG:DATA}. The HR-RBW-filtering condition for the data
shown in figure~\ref{FIG:DATA} removes frequency points with power
levels higher than 0.1 from the exponential fit shown in
figure~\ref{FIG:DATA} (c). A constant fit shown in
figure~\ref{FIG:DATA} (b) was used for the
LR-RBW-filtering. Figure~\ref{FIG:DATAFILTER} shows the normalized
power excess distributions after applying the LR-RBW-filtering only
(left) and the two-staged filtering (right), respectively. Since the
latter figure has more data points (entries), it has filtered out less
data and thus it is more efficient. Here, the difference is not that
significant in the absence of the axion signal and when no further
analysis processing is done, while we are able to demonstrate the
effect is significant in the presence of the signal and with the full
analysis procedure in the following sections.

\subsection{Standard halo}\label{SEC:HALO}
The signal window of the standard halo axion dark matter is about
4~kHz for an axion frequency of about 1~GHz as in
refs.~\cite{12TB-PRL,12TB-PRX}.
\begin{figure}[h]
  \centering
  \includegraphics[width=0.90\columnwidth]{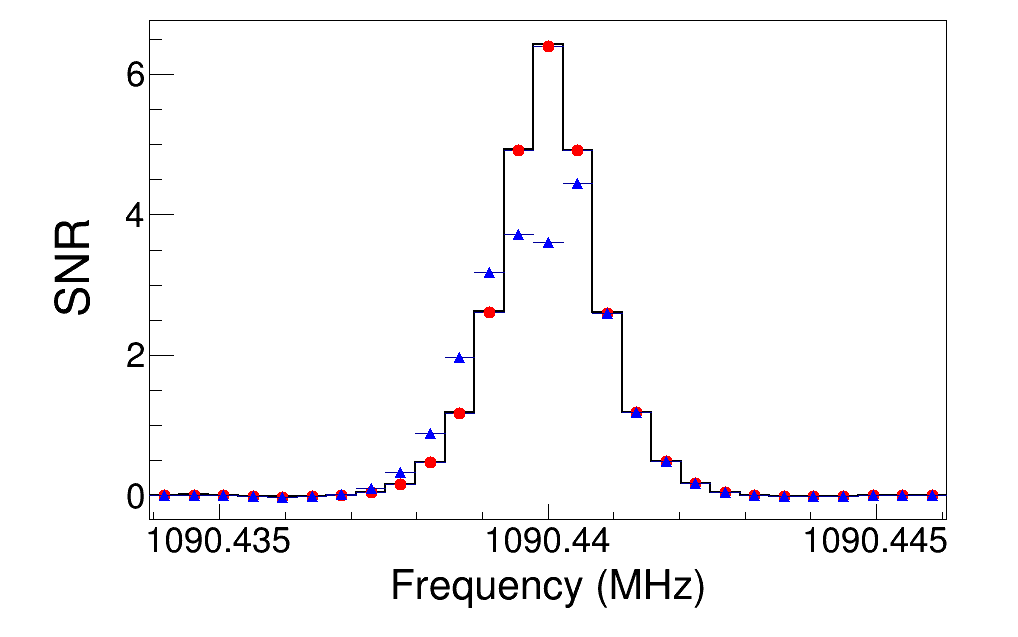}
  \caption{SNRs as a function of frequency around the signal region
    from 10000 simulated experiments. The black line is the case with
    an $\epsilon_{\rm SNR}$ of 100\%. The red circles are with
    applying both the HR-RBW-filtering and the LR-RBW-filtering and
    the blue triangles with the LR-RBW-filtering only.}  
  \label{FIG:SNRofHALO}
\end{figure}
The standard halo axion dark matter signal was reconstructed following
not only the procedure in ref.~\cite{12TB-PRL} which includes the
LR-RBW-filtering, but also the additional HR-RBW-filtering.
Hence the procedure steps in sequence, in view of
figure~\ref{FIG:SIMUL}, are the HR-RBW-filtering to the individual
power spectra whose RBW is 0.025~Hz (figure~\ref{FIG:SIMUL} (a)),
averaging over the 13 power spectra (figure~\ref{FIG:SIMUL} (b)),
merging 400 frequency points resulting in an RBW of
10~Hz (figure~\ref{FIG:SIMUL} (c)), and finally background subtraction
(figure~\ref{FIG:SIMUL} (f)).
The LR-RBW-filtering was applied to the
data shown in figure~\ref{FIG:SIMUL} (f). After the relevant
overlapping and coadding steps, the SNRs as a function of frequency
around the signal region from 10000 simulated experiments are shown in
figure~\ref{FIG:SNRofHALO}.
The red circles and blue triangles in figure~\ref{FIG:SNRofHALO} are
from data with the aforementioned narrow peak in the signal region,
where the former was gone through the HR-RBW-filtering and
LR-RBW-filtering and the latter the LR-RBW-filtering only.
The black line in figure~\ref{FIG:SNRofHALO} is from data without the
narrow peak and its efficiency is 100\% since it does not have any
filtering.
As shown in figure~\ref{FIG:SNRofHALO}, the black line and red circles
are almost identical, which means the HR-RBW-filtering efficiency is
almost 100\%.
The SNR represented by the blue triangles, however, are significantly
degraded at the signal frequency compared to that by the red circles,
because the LR-RBW-filtering was triggered for narrow peaks in the
blue triangle data, but not to the red circle data which has already
removed such peaks with the HR-RBW-filtering.
Such an SNR degradation is attributed by the LR-RBW-filtering
condition, because the filtering usually removes the neighboring
points as well as the narrow peak. The blue triangles resulted from
removing $\pm5$ data points with respect to the peak position as done
in ref.~\cite{12TB-PRL}.
\begin{figure}[h]
  \centering
  \includegraphics[width=0.95\columnwidth]{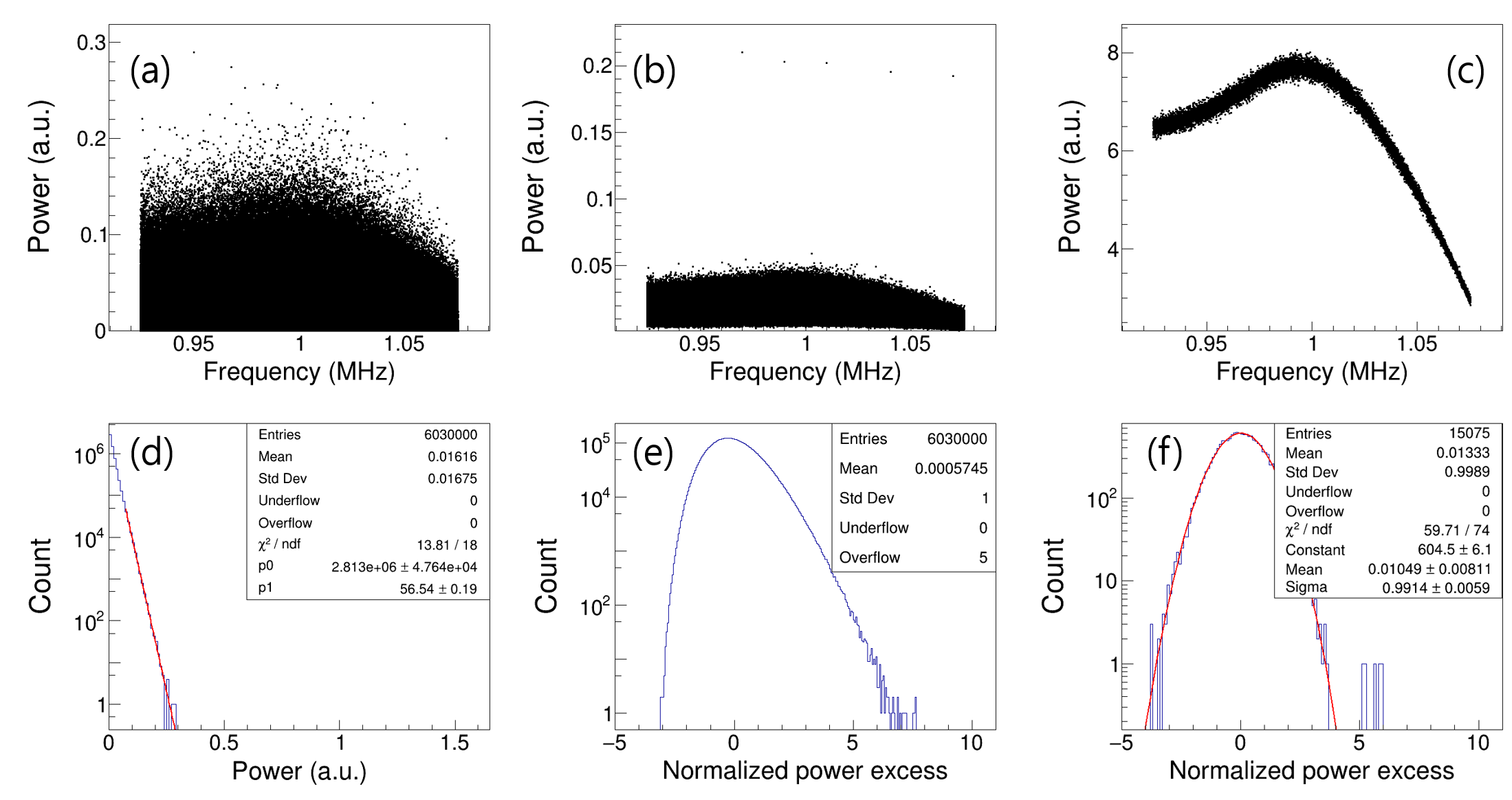}
  \caption{The same as figure~\ref{FIG:SIMUL} except for the frequency
    of the cavity mode, where it is 40~kHz away from the axion
    mass. Hence the axion signal is positioned at frequencies around
    the 4th narrow peak clearly shown in (b).
    The 5 overflows in (e) correspond to the 5 excesses whose SNRs are
    larger than 5 in (f).}
  \label{FIG:SIMUL4}
\end{figure}
There exist subtle differences between the black line and the red
circles in figure~\ref{FIG:SNRofHALO} due to the narrow peak in the
signal region when the frequency of the cavity mode sits away from the
axion mass as shown in figure~\ref{FIG:SIMUL4} (b).
The data shown in figure~\ref{FIG:SIMUL4} is the same as
figure~\ref{FIG:SIMUL} except for the frequency of the cavity mode,
where it is 40~kHz away from the axion mass.
Due to the cavity response profiles, the amplitude of all the narrow
peaks are suppressed down to below 0.35 as shown in
figure~\ref{FIG:SIMUL4} (a) and (d), which caused the HR-RBW-filtering
to overlook these points.
The LR-RBW-filtering was applied to the signal region for this data
which still retains the spurious peaks as seen in
figure~\ref{FIG:SIMUL4} (f),
but the amplitude of the signal power was also suppressed by the cavity
response profiles, which resulted in the negligible filtering effect
to the SNR as shown in red circles in figure~\ref{FIG:SNRofHALO}. This
subtle effect is more visible in the tidal stream axion dark matter
search in section~\ref{SEC:TIDAL}.
For tuning steps where the cavity mode’s resonance frequency sits
farther away from the axion mass, the spurious narrow peak will not be
as much above the noise and can be overlooked by both the
HR-RBW-filtering and LR-RBW-filtering. Only the surviving spurious
peaks in the signal region from each frequency tuning step participate
in the overlapping process. Depending on the filtering conditions, the
peaks' contribution to the SNR during overlapping can vary.
For LR-RBW-filtering, most of the data at the signal frequency is
removed alongside the spurious peak when the cavity resonance is
closer to the signal frequency. On the other hand, HR-RBW-filtering
can do ``pinpoint removals'' of spurious peaks without affecting the
rest of the data at high resolution. Thus, the contribution of
surviving spurious peaks is higher for the blue triangles before
coadding, due to the absence of the data when the cavity resonances
are close to the axion mass.
This leads to a higher-than-usual power excess at the signal
frequency, at the cost of retained data at that point.
The combined effects result in the SNR being skewed to the left, with
some points showing better SNRs compared to the black line, after
coadding with an asymmetric, boosted Maxwellian axion signal shape.
In contrast, the red circles has relatively more data coming from the
axion signal, and the surviving spurious peaks do not have such a
pronounced effect on the SNR. As a result, the final coadded SNR
closely follows the black line.

The SNR of about 6 by the red circle in figure~\ref{FIG:SNRofHALO}
would be close to 5 after considering the aforementioned background
subtraction efficiency and additional systematic effects, e.g., noise
calibrations. Hence applying a typical threshold of 3.718 of the
normalized power excess, one can get a one-sided 90\% upper limit
corresponding to an SNR of 5.
Since the parameters used are similar to those in
ref.~\cite{12TB-PRL}, the sensitivity from the data generated in this
work would be similar to that achieved in ref.~\cite{12TB-PRL}, i.e.,
DFSZ sensitivity at a 90\% CL.
\subsection{Tidal stream}\label{SEC:TIDAL}
The tidal stream axion dark matter from the Sagittarius dwarf
galaxy~\cite{TIDAL1,TIDAL2} would have a signal window of about
150~Hz for an axion frequency of about 1~GHz as in
ref.~\cite{12TB-PRD}.
Even with such a narrow signal window, the averaging and overlapping
of power spectra can be applied to the
experiment~\cite{12TB-PRL,12TB-PRX} to increase the SNR, according to
ref.~\cite{12TB-PRD}.
\begin{figure}[h]
  \centering
  \includegraphics[width=1.0\columnwidth]{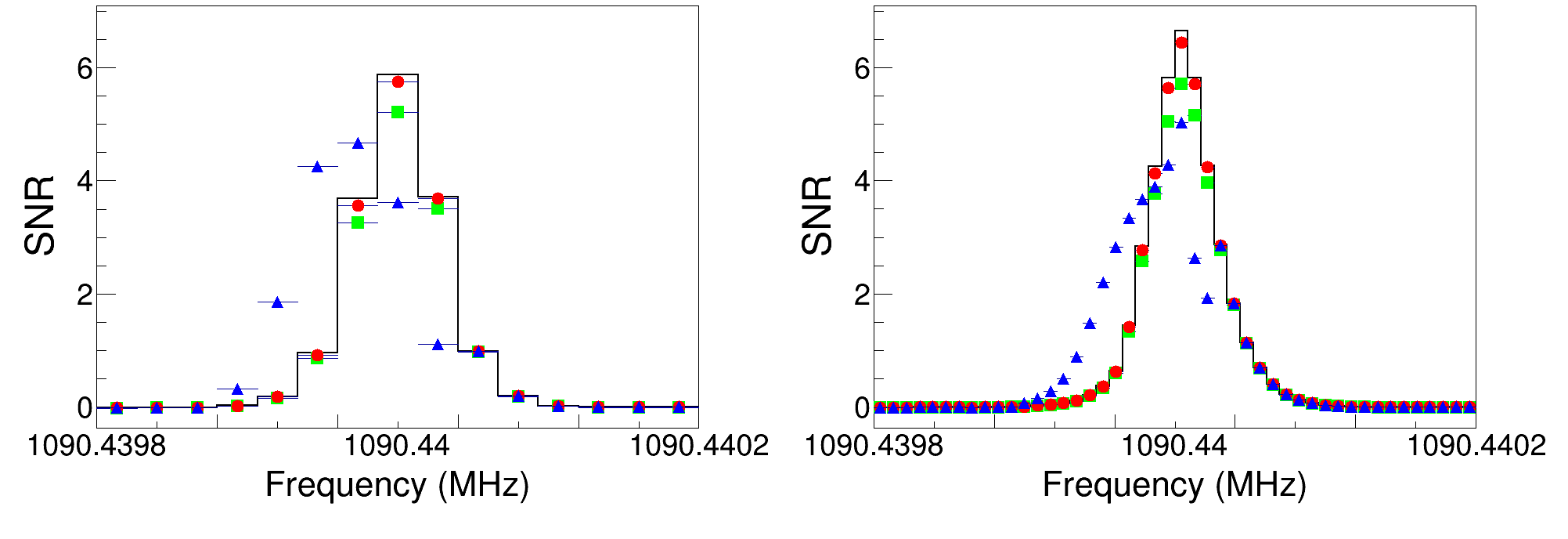}
  \caption{SNRs as a function of frequency around the signal region
    from 10000 simulated experiments. The black line is the case with
    an $\epsilon_{\rm SNR}$ of 100\%. The red circles are with
    applying both the HR-RBW-filtering and the LR-RBW-filtering and
    the blue triangles with the LR-RBW-filtering only, where the
    LR-RBW-filtering condition is the looser one, i.e., the normalized
    power excesses greater than 4.5.
    The green rectangles are with applying both filtering but with
    the tighter LR-RBW-filtering with the normalized power excesses
    greater than 3.5.    
    Left followed the RBW reduction in ref.~\cite{12TB-PRD}, i.e.,
    10~Hz to 30~Hz, then applied the coadding of 5 frequency points to
    ensure the signal window of 150~Hz, while right maintains the RBW
    of 10~Hz, then applied the coadding of 15 frequency points.}  
  \label{FIG:SNRofTIDAL}
\end{figure}
Hence the analysis procedure is basically the same as that for the
standard halo axion dark matter search in the previous
section~\ref{SEC:HALO}.
Due to the absence of a dedicated rescan schedule in
ref.~\cite{12TB-PRD}, a tighter LR-RBW-filtering, i.e., the removal
of normalized power excesses higher than 3.5, was applied.
The neighboring $\pm1$ data points with respect to the peak position
were also removed. We can in general relieve the condition to 4.5 from
3.5 assuming a rescan. Allowing the averaging and overlapping and
after the relevant coadding procedure, the SNRs as a function of
frequency around the signal region from 10000 simulated experiments
are shown in the left plot of figure~\ref{FIG:SNRofTIDAL}.
The green rectangles in left plot of figure~\ref{FIG:SNRofTIDAL} were
obtained with the same LR-RBW-filtering condition in
ref.~\cite{12TB-PRD} and the SNR at the signal frequency is about 91\%
with respect to that of the red circles in the same plot, which is
consistent with the LR-RBW-filtering efficiency in
ref.~\cite{12TB-PRD}. The red circles were obtained with the loose
LR-RBW-filtering assuming the rescan.
The SNR can be further improved from a coadding procedure with higher
resolution RBW as seen in ref.~\cite{12TB-PRL}. The left and right
plots of figure~\ref{FIG:SNRofTIDAL} result from 5 coadded frequency
points with 30~Hz intervals and 15 coadded frequency points with 10~Hz
intervals, respectively. This ensures that both cases cover the signal
window of 150~Hz.
The SNRs in figure~\ref{FIG:SNRofTIDAL} were achieved with DFSZ axion
dark matter that constitutes about 19.2\% of the local dark matter
density and the SNR by the red circle at the signal frequency in right
plot is similar to the SNR of the standard halo axion dark matter
search assuming $\rho_a=0.45$~GeV/cm$^3$.
Therefore, a DAQ sensitive to the standard halo DFSZ axion dark
matter that makes up 100\% of the local dark matter density can also
search for tidal stream DFSZ axion dark matter that constitutes about
19.2\% of the local dark matter density simultaneously.
Without HR-RBW-filtering, the SNRs by the blue triangles in
figure~\ref{FIG:SNRofTIDAL} are worse at the signal frequency due to
the same reason discussed in section~\ref{SEC:HALO}.

The aforementioned discrepancy in the SNR of the black line and the
red circles being more noticeable here than compared to
figure~\ref{FIG:SNRofHALO} (for the standard halo axion dark matter
search) is due to different LR-RBW-filtering conditions in this
work. The tidal stream filter removes 30~Hz width of frequency points
creating a larger effect on the 150~Hz signal window, compared to the
standard halo, where the filter removes 110~Hz for a model whose
signal window is 4050~Hz.
\subsection{Big flow}
\begin{figure}[h]
  \centering
  \includegraphics[width=1.0\columnwidth]{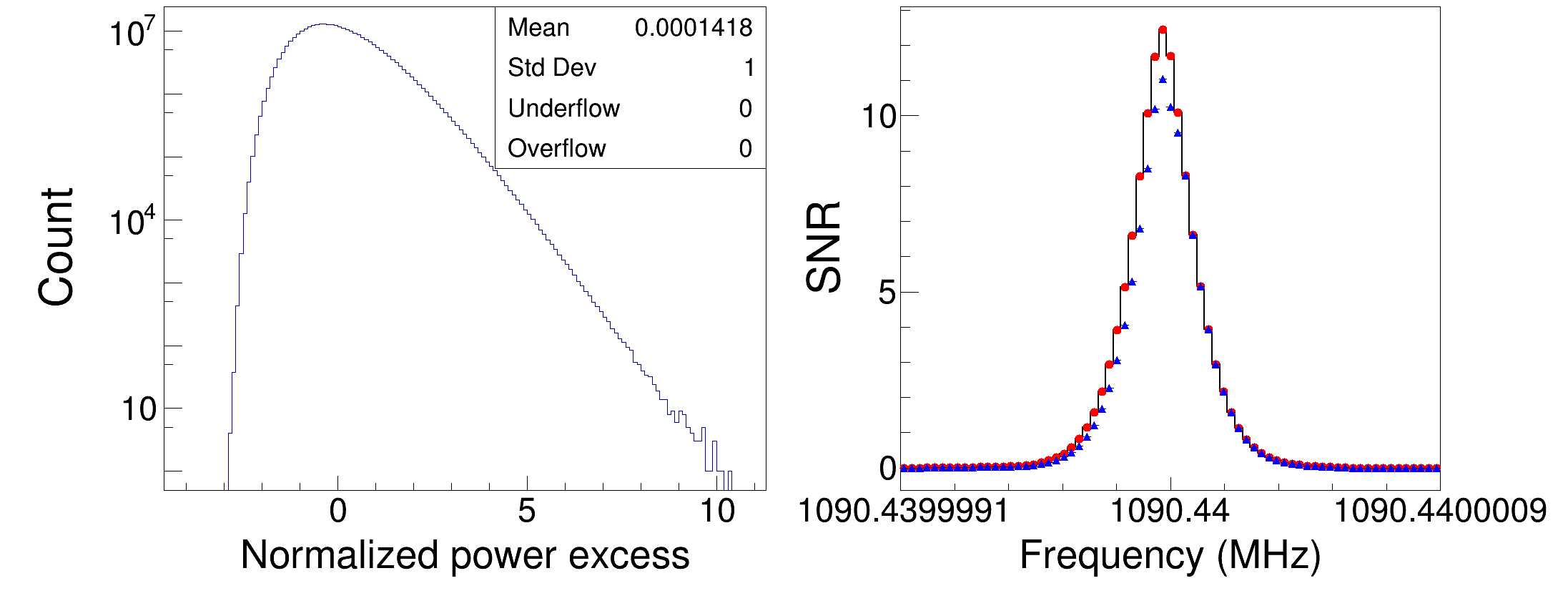}
  \caption{The normalized power excess distribution from the
    background only data after coadding 25 frequency points (left) and
    the SNRs as a function of frequency around the signal region from
    10000 simulated experiments (right).
    The black line in the right plot is the case with an
    $\epsilon_{\rm SNR}=100$\%. The blue triangles and red circles are
    from data with and without the narrow peak background in the
    signal region, respectively, where they are with the
    HR-RBW-filtering.}  
  \label{FIG:SNRofBIG}
\end{figure}

The signal width of the big flow axion dark matter whose mass is about
1~GHz is expected to be less than 1~Hz. For such an extremely narrow
signal window, the averaging and overlapping of power spectra are not
applicable due to the signal frequency shift resulting from the
rotational and orbital motions of Earth~\cite{12TB-PRD}. Therefore,
only one power spectrum out of 13 power spectra collected in a
frequency step was used for the big flow axion dark matter search.
In order to benefit from the cavity resonance profiles and avoid the
overlapping of the power spectra, we only took a $\pm5$~kHz window
with respect to the IF in individual power spectra for a given
frequency tuning step of 10~kHz. Then, all the power spectra from each
resonance frequency were combined without the averaging and
overlapping to build the grand power
spectrum. After coadding 25 frequency points to ensure a signal window
of 0.625~Hz, figure~\ref{FIG:SNRofBIG} shows the normalized power
excess distribution from the background only data (left) and the SNRs
as a function of frequency around the signal region from 10000
simulated experiments (right).
As shown in the left plot of figure~\ref{FIG:SNRofBIG}, the background
in this case follows a non-gaussian distribution. From this
distribution, we can get a one-sided 90\% upper limit corresponding to
an SNR of 10 by a threshold of 8.827 of the normalized power excess.
Considering the inefficiencies and systematic effects, an SNR of about
12 marked as the red circle in the right plot of
figure~\ref{FIG:SNRofBIG} would conservatively approach 10.
The blue triangles in the right plot of figure~\ref{FIG:SNRofBIG} are
from data with the narrow peaking background in the signal region,
thus showing unavoidable SNR degradation compared to the red circles
from data without the peaking.

The SNR of the red circle at the signal frequency in the right plot of
figure~\ref{FIG:SNRofBIG} was achieved with DFSZ axion dark matter
that constitutes about 12.4\% of the local dark matter density.
Therefore, big flows with these characteristics can be searched for
using the same DAQ that is sensitive to standard halo DFSZ axion dark
matter that constitutes 100\% of the local dark matter density
$\rho_a=0.45$~GeV/cm$^3$.

\section{Summary}
We show axion dark matter searches covering the standard halo, the
tidal stream, and the big flow employing a DAQ channel with a single
HR RBW, without sacrificing the $\epsilon_{\rm DAQ}$, where the DAQ
process includes online FFTs and writing the outputs to disk.
Assuming that the total amount of data from DAQ is sensitive to the
standard halo DFSZ axion dark matter whose $\rho_a=0.45$~GeV/cm$^3$,
then the same data also provides the tidal stream and the big flow
axion dark matter searches that would be sensitive to DFSZ axion dark
matter that constitute 19.2\% and 12.4\% of the local dark matter
densities, respectively, at a 90\% CL. We also report that the
filtering of the individual power spectra acquired with an HR RBW,
e.g., for the big flow search can prevent a possible significant
degradation in the SNR from the searches in the lower resolution
RBW's, i.e., the standard halo and tidal stream searches.
The difference in the degradation of SNR is quantified in
table~\ref{Final_comp}. Table~\ref{Final_comp} provides the expected
SNRs assuming DFSZ axions make up 100\% of the local dark matter
density and shows the novelty of this work relative to
refs.~\cite{12TB-PRL,12TB-PRD}. The HR RBW data acquisition allows for
big flow searches to be possible and better SNR using two-staged
filtering in SHM and tidal searches compared to previous analyses.
\begin{table} [h]
  \centering
  \begin{tabular}{ | l | c | c |} \hline
& \cite{12TB-PRL,12TB-PRD} & this work \\ 
& with the LR-RBW-filtering only & with the two-staged filtering \\ \hline\hline  
 SHM          &4.4 & 6.4    \\ \hline  
tidal stream  &4.7 & 5.7    \\ \hline  
  \end{tabular}
  \caption{Expected SNRs assuming DFSZ axions with
    $\rho_a=0.45$~GeV/cm$^{3}$, when a narrow spurious peak is located
    at the signal region. The numbers with the two-staged filtering
    were chosen at the axion mass, while those with LR-RBW-filtering
    only as the maximum value around the axion mass in their skewed
    SNR distributions shown in figures~\ref{FIG:SNRofHALO} and
    \ref{FIG:SNRofTIDAL}.}
  \label{Final_comp}
\end{table}

\acknowledgments
This work was supported by a Korea University Grant, the National
Research Foundation of Korea (NRF) grants funded by the Korea
government (MSIT) (RS-2025-00556247) and (RS-2022-00143178), and the
Korea Basic Science Institute (National research Facilities and
Equipment Center) grant funded by the Korea government (MSIT)
(NFEC-2019R1A6C1010027).

\end{document}